# PERFORMANCE OF A DOWNCONVERTER TEST-ELECTRONICS WITH MTCA-BASED DIGITIZERS FOR BEAM POSITION MONITORING IN 3.9GHZ ACCELERATING CAVITIES

T. Wamsat*†, N. Baboi†, B. Lorbeer†, P. Zhang†‡
†Deutsches Elektronen-Synchrotron DESY, Hamburg, Germany
‡The University of Manchester, Manchester, U.K.

## *Abstract*

Beam-excited higher order modes (HOM) in 3.9GHz accelerating cavities are planned to be used for beam position monitoring at the European XFEL. The selected HOMs are located around 5440MHz, with a bandwidth of 100MHz and 9060MHz, with a bandwidth of 50MHz. A downconverter electronics, built for tests at FLASH, converts the HOMs to an intermediate frequency of 70MHz. The MTCA (Micro Telecommunications Computing Architecture) standard will be used for the XFEL. Thus it is important to have a performance study of the downconverter test-electronics using the MTCA digitizer card SIS8300. In the digitizer the IF frequency of 70MHz is undersampled with a clock frequency of 108MS/s. The paper presents the performance of the digitizer together with the test-electronics. A comparison with a 216MS/s VME (Versa Module Eurocard) digitizer is made.

## INTRODUCTION

The ACC39 module in FLASH, DESY [1] contains four third harmonic cavities operating at 3.9GHz as shown in Fig. 1. An electron bunch passing through these cavities generates wakefields, which can be decomposed into higher order modes (HOM). These HOMs, available from existing coupler, can be used to determine transverse beam positions in the cavities [2] [3].

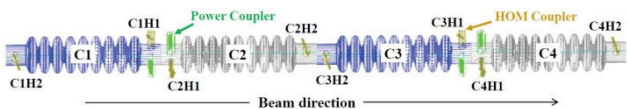

Figure 1: Schematic of the four cavities within ACC39.

In 3.9GHz cavities HOMs around 5440MHz and 9060MHz are used to determine the beam position [4].

### *Measurement setup*

The HOM signals are downconverted to an intermediate frequency (IF) of 70MHz which is then digitized. Fig. 2 shows the schematic block diagram of the downconverter, built by FNAL [4].

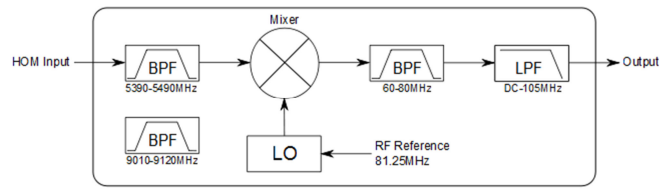

Figure 2: Schematics of the downconverter analog box.

A bandpass filter (BPF) with an appropriate frequency band filters the HOM signal and then it is connected to a mixer. A local oscillator (LO) which uses a 81.25MHz reference from the master oscillator of FLASH generates the required LO frequency. The IF signal is then filtered with a 20MHz bandpass filter. After a proper amplification, the IF is further filtered by a lowpass filter (LPF) to reduce the noise. To switch the analog box to another frequency, the bandpass filter was swapped and the frequency of the LO was changed accordingly to ensure an IF of 70MHz.
The analog electronics is shown in Fig. 3.

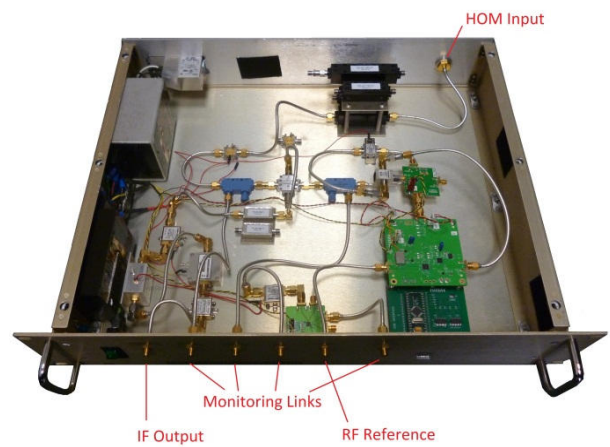

Figure 3: Downconverter analog box.

The downconverted HOM signal was digitized by the VME and also by the MTCA. The former used a sampling rate? of 216 MHz, and the second 108MHz. The setup used to make a comparison between the VME and MTCA is shown in Fig. 4.

___

*thomas.wamsat@desy.de

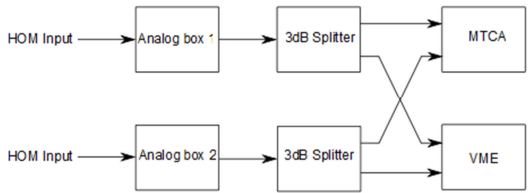

Figure 4: Set-up with VME and MTCA.

An example of the IF output signal of the downconverter is shown in Fig. 5.

The MTCA system is described in detail in ref. [6].

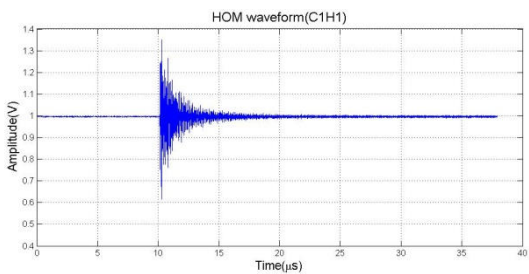

Figure 5: Downconverted HOM signal with MTCA.

*Measurement scheme*

The schematic of the measurement setup is shown in Fig. 6. An electron bunch of approximately 0.5nC is passing the ACC39 module. Two steering magnets located in front of ACC1 are used to produce horizontal and vertical offsets of the electron bunch in ACC39. Two beam position monitors (BPM-A and BPM-B) are used to record transverse beam position before and after ACC39. A straight line trajectory of the bunch is produced between those two BPMs by switching off the accelerating field in ACC39 and all quadruples close to it.

Therefore, the transverse offset of the electron bunch in each cavity can be determined by interpolating the readouts of the two BMPs, as shown in Fig. 7.

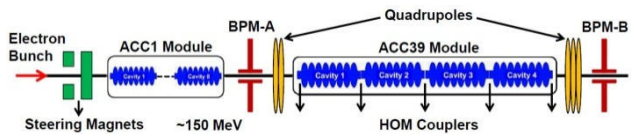

Figure 6: Schematic of measurement setup (not to scale, cavities in ACC1 are approximately three times larger than those in ACC39).

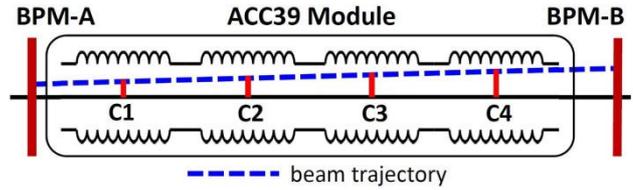

Figure 7: Position interpolation into the center of each cavity.

# BEAM POSITION DETERMINATION

To determine the beam position in each cavity a procedure based on Singular Value Decomposition (SVD) is used and the details are described in ref. [5].

At the beginning of the measurement a calibration was made. The beam was moved in a 2D grid manner by changing the steerer current. The resultant beam positions at BPM-A and BPM-B are shown in Fig. 8 and Fig. 9.

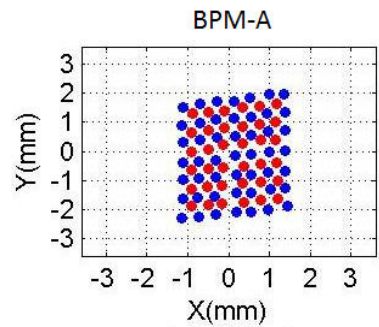

Figure 8: The readouts of BPM-A during the 2D grid scan (calibration: blue, validation: red).

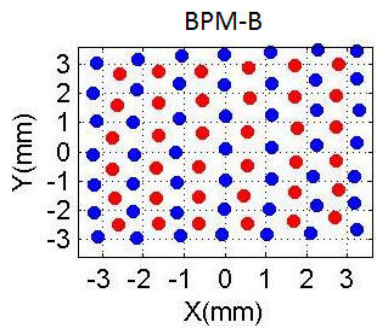

Figure 9: The readouts of BPM-B during the 2D grid scan (calibration: blue, validation: red).

The set of beam positions marked in blue were used for calibration of the HOM signals into beam offsets. The ones in red were applied this calibration and were used to estimate the system performance.

A result of the beam position prediction for calibration and validation using the HOM signal with SVD compared to the interpolated beam position (x,y) using BPM-A and BPM-B are shown in Fig. 10 and Fig. 11.

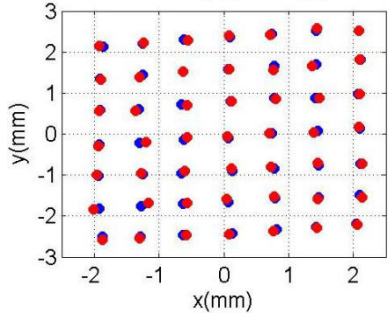

Figure 10: Measurement (blue) and prediction (red) of beam position from calibration.

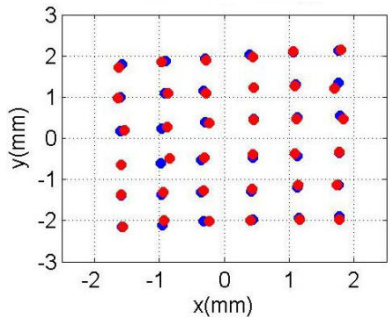

Figure 11: Measurement (blue) and prediction (red) of beam positions from validation.

The difference between measured and predicted transverse beam position from calibration and validation is shown in Fig. 12.

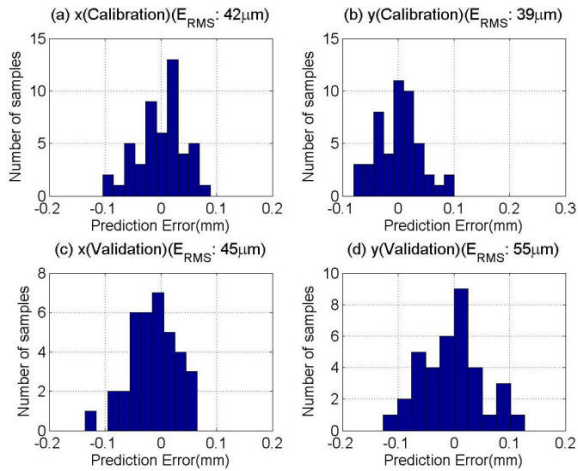

Figure 12: Difference of measured and predicted beam positions.

Note that the denoted resolution in the plots is in a global sense, since the RMS is calculated on the prediction errors for all different beam offsets. It is not the resolution which is usually used in beam diagnostics for certain positions. This will be clarified in the next section.

## RESOLUTION ESTIMATION

For this measurement couplers C1H1, C2H1 and C3H1 were used (see Fig. 1). To estimate the resolution a hundred beam pulses were taken for each of several positions shown in Fig. 13 as measured with BPM-B. In this measurement only the MTCA digitizer was used.

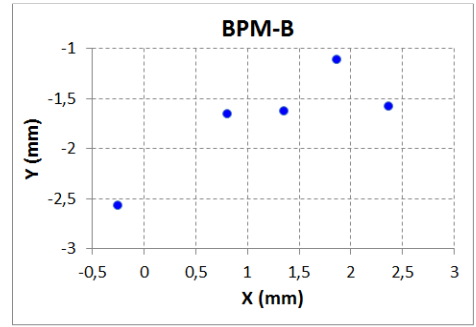

Figure 13: Selected beam positions for resolution studies.

Fig. 14 shows the analysis of one measurement. Fig. 14(a) shows the calibration with the position determination from the BPM in blue and the prediction from the SVD in red. In Fig. 14(b) all 100 beam pulses are painted within it and the deviation between these hundred beam pulses can be obtained. Fig. 14(c) and (d) show the resultant resolution for x and y direction at a specific position.

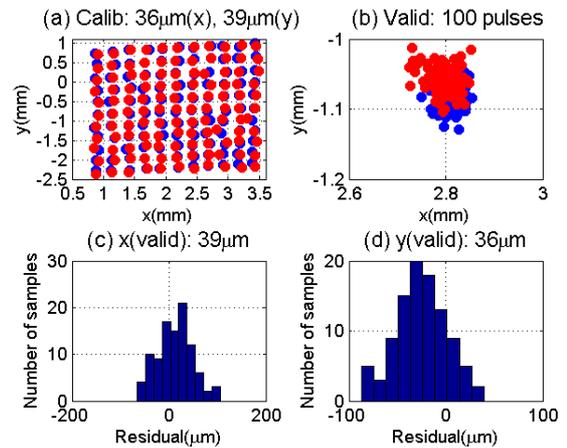

Figure 14: Analysis of the 100 beam pulses taken from coupler C3H1 at 9060MHz with MTCA at BPM-B position (2.37, -1.58), see Fig. 13.

In Fig. 15 the measured and predicted beam positions for the same position, three cavities are shown.

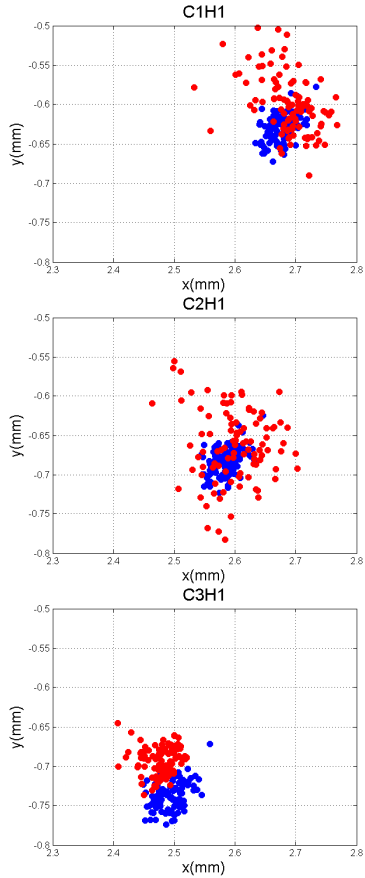

Figure 15: Measured (blue) and predicted (red) beam positions in the three cavities C1 to C3.

Fig. 16 to Fig. 18 show the resolution results taken with a hundred beam pulses with the MTCA for the selected beam positions shown in Fig. 13. The dashed lines indicate the MTCA resolution in a global sense using a cross-validation technique [5]. The solid lines are results from the VME digitizers.

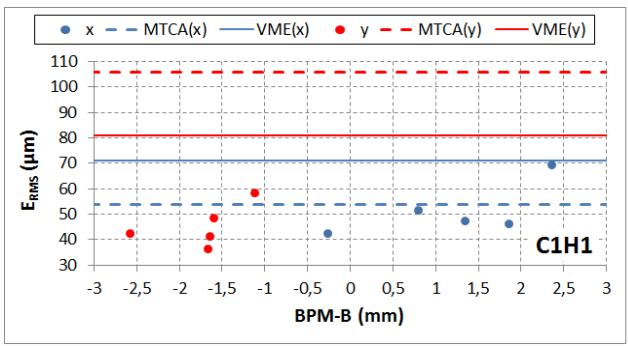

Figure 16: Position resolution of x and y in C1.

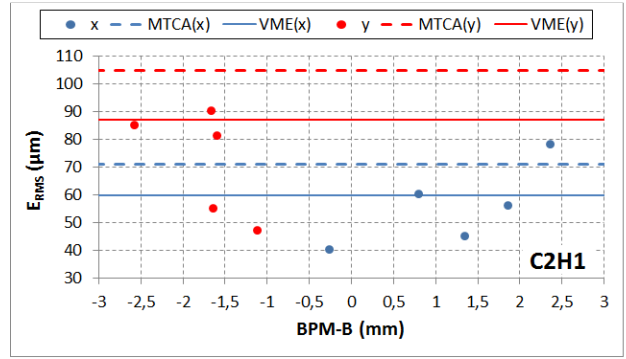

Figure 17: Position resolution of x and y in C2.

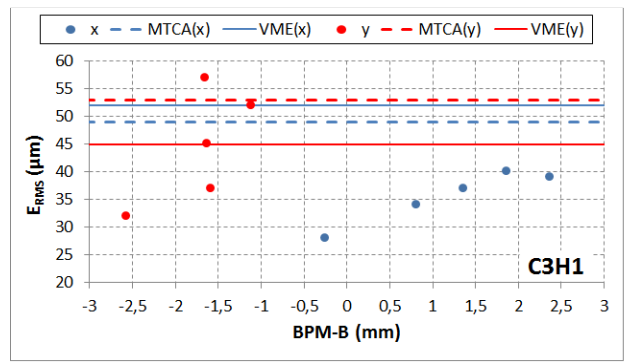

Figure 18: Position resolution of x and y in C3..

The reached resolution with both digitizers is similar. But it has to be considered that the VME digitizer works with 216MS/s and the MTCA with 108MS/s which means an undersampling of the used 70MHz. The resolution obtained for several beam positions is in general better than the global resolution.

## CONCLUSIONS

The MTCA SIS8300 digitizer card is a reasonable alternative to the VME digitizer. The resolution in this application is nearby the same although the sampling rate of the MTCA causes an undersampling.

## OUTLOOK

A downconverter with an IF of 30MHz is currently built by FNAL in order to avoid undersampling. This will be installed at FLASH. A MTCA-based digitizer card with a higher sampling rate (500Ms/s), under development at DESY, is considered for the XFEL.